\documentclass[epj]{svjour}
\usepackage{latexsym}
\usepackage{graphics}
\usepackage{epsfig}
\usepackage{amsfonts}
\begin{document}
\newcommand{\be}{\begin{equation}}
\newcommand{\ee}{\end{equation}}
\newcommand{\La}{\cal{L}}
\newcommand{\no}{\noindent}
\newcommand{\ba}{\begin{eqnarray}}
\newcommand{\ea}{\end{eqnarray}}
\newcommand*{\pd}{\partial}
\newcommand*{\pdm}{\pd_{\mu}}
\newcommand*{\pdn}{\pd_{\nu}}
\newcommand*{\pdi}{\pd_{i}}
\newcommand*{\pda}[1]{\pd_#1}
\newcommand*{\bea}{\begin{eqnarray}}
\newcommand*{\eea}{\end{eqnarray}}
\newcommand*{\pref}[1]{(\ref{#1})}
\newcommand*{\img}{\mathrm{im}}
\newcommand*{\mn}{{\mu\nu}}
\newcommand*{\rr}{\mathbb{R}}
\newcommand*{\gh}{\mathrm{gh}}
\newcommand*{\sdet}{\mathrm{sdet}}
\newcommand*{\str}{\mathrm{str}}
\newcommand*{\uth}{^\mathrm{th}}
\newcommand*{\ust}{^\mathrm{st}}    
\newcommand*{\prefr}[2]{(\ref{#1}-\ref{#2})} 
\newcommand*{\piclinecol}{red }
\title{The High-Temperature Phase of Landau-Gauge Yang-Mills Theory}
\author{Axel Maas\inst{1} \and Jochen Wambach\inst{1,2} \and Reinhard Alkofer \inst{3}}     
\institute{Gesellschaft f\"ur Schwerionenforschung mbH, Planckstra{\ss}e 1, 
D-64291 Darmstadt, Germany \and Institute for Nuclear Physics, Darmstadt University of Technology, 
Schlo{\ss}gartenstra{\ss}e 
9, D-64289 Darmstadt, Germany \and Institute of Physics, University of Graz, Universit\"atsplatz 5, A-8010 Graz, 
Austria}
\date{Received: date / Revised version: date}
\abstract{
The properties of the high-temperature phase of Yang-Mills theory in Landau gauge are investigated by extending an 
earlier study on the infinite-temperature limit to finite temperatures. To this end the Dyson-Schwinger equations for 
the propagators of the gluon and the Faddeev-Popov ghost are solved analytically in the infrared and numerically at 
non-vanishing momenta. Gluons, polarized transversely with respect to the heat bath are found to comply with the 
Gribov-Zwanziger and the Kugo-Ojima scenario, while longitudinally polarized gluons are screened. Therefore the 
high-temperature phase is strongly interacting. It is furthermore conjectured that Yang-Mills theory undergoes a 
first-order phase transition. Indications are found that at high temperatures the thermodynamic properties are nearly 
those of an ideal gas, although long-range interactions prevail.
\PACS{
      {11.10.Wx}{Finite-temperature field theory} \and
      {11.15.-q}{Gauge field theories} \and
      {12.38.-t}{Quantum chromodynamics} \and
      {12.38.Aw}{General properties of QCD} \and
      {12.38.Lg}{Other nonperturbative calculations} \and
      {12.38.Mh}{Quark-gluon plasma} \and 
      {14.70.Dj}{Gluons}
     } 
} 
\maketitle
\section{Introduction}\label{sintro}

Many properties of the high-temperature phase of QCD and even of pure Yang-Mills theory are not yet understood, especially the infrared behavior and thus the fate of confinement. The equation 
of state, as found in lattice calculations, exhibits an almost trivial Stephan-Boltzmann behavior of the ideal gas \cite{Karsch:2003jg}. This is also found in so-called weak-coupling expansions \cite{Kajantie:2000iz}. On the other hand, it has already long ago been argued that the microscopic processes cannot be purely perturbative \cite{Linde:1980ts}. This is exemplified by the non-vanishing spatial string-tension \cite{Manousakis:1986jh}. In addition, in the limit of infinite temperature, at least part of the gluon spectrum shows confinement \cite{Maas:2004se} in accordance with the scenarios of Kugo and Ojima \cite{Kugo:gm} and of Gribov and Zwanziger \cite{Gribov:1977wm,Zwanziger:2003cf}. Also at finite temperature indications of such effects have been seen using various techniques \cite{Cucchieri:2001tw,Cucchieri:2000cy,Zahed:1999tg}. This clearly demonstrates that nonperturbative quantum effects persist in that limit.

The objective here is to understand the properties of the gluons and Faddeev-Popov ghosts, in particular in the 
infrared, in the high-temperature phase. This work is therefore an extension of studies at infinite temperature 
\cite{Maas:2004se} and complements studies at temperatures below the phase transition \cite{Gruter:2004bb}.

To accomplish this task, the Dyson-Schwinger equations (DSEs) \cite{Dyson:1949ha} for the propagators of the gluon and 
the Faddeev-Popov ghost are investigated. A clear separation of soft and hard degrees of freedom is found. Indications 
will be discussed that this separation of scales is the reason why a microscopically highly non-trivial and especially 
nonperturbative theory can exhibit quite simple macroscopic properties as is the case for the equation of state.

The paper is organized as follows. In section \ref{saspects} Yang-Mills theory and signals of confinement are briefly 
reviewed. The Dyson-Schwinger equations are introduced in section \ref{sdse}. An analytical treatment of the infrared 
sector is given in section \ref{sir}. The consequences of the necessary truncations and renormalization will be 
discussed in section \ref{sstrunc}. Numerical solutions beyond the infrared will be presented in section \ref{snum}. 
This includes comments on the thermodynamic potential and Schwinger functions. A discussion of the results and the 
phase structure are finally given in section \ref{sdiscuss}. A summary and concluding remarks close the paper in 
section \ref{ssummary}. Some technical issues are deferred to two appendices.

\section{Aspects of Yang-Mills theory}\label{saspects}

The following investigations are restricted to pure Yang-Mills theory as substantial evidence exists that the 
nonperturbative features of QCD are generated in the gauge sector. Hence, the theory studied here is an equilibrium 
Yang-Mills theory governed by the Euclidean Lagrangian\footnote{The hermiticity assignment for the ghosts is different 
from the conventional, although valid for Landau gauge \cite{Alkofer:2000wg}.} \cite{Alkofer:2000wg,Kapusta:tk}
\bea
\mathcal{L}&=&\frac{1}{4}F_{\mu\nu}^aF_{\mu\nu}^a+\bar c^a \pdm D_\mu^{ab} c^b\nonumber\\
F^a_{\mu\nu}&=&\pdm A_\nu^a-\pdn A_\mu^a-gf^{abc}A_\mu^bA_\nu^c\nonumber,\\
D_\mu^{ab}&=&\delta^{ab}\pdm+gf^{abc}A_\mu^c\nonumber~.
\eea
\no Here $F_{\mu\nu}^a$ denotes the field strength tensor, $D_\mu^{ab}$ the covariant derivative, $g$ the gauge 
coupling and $f^{abc}$ the structure constant of the gauge group, which is SU(3) unless otherwise noted. $A_\mu^a$ 
denotes the gluon field, and $\bar c^a$ and $c^a$  are the Faddeev-Popov ghost fields describing part of the quantum 
fluctuations of the gluon field. We choose the  Landau gauge, which for technical reasons is best suited for the 
purpose at hand \cite{Alkofer:2000wg}.

The main focus of the present work is the fate of confinement at temperatures above the phase transition. It is 
therefore necessary to be able to extract information about confinement. Part of this information is encoded in 
the pertinent 2-point functions. The corresponding signals will be only listed here. For a brief introduction to 
confinement in covariant gauges see ref.\ \cite{Schleifenbaum:2004mm}.

A sufficient criterion for the presence of confinement is based on the fact that no K\"all{\'e}n-Lehmann 
representation of a particle exists if its spectral function is not positive semi-definite. It is then not part of the 
physical spectrum and thus confined, c.f.\ \cite{Oehme:bj}. This is the case if the corresponding propagator $D$ 
vanishes at zero momentum,
\be 
\lim_{p^2 \to 0}D(p^2)=0.\label{Oehme} 
\ee
\no This is essentially the behavior expected for the propagator of a confined gluon.

A second criterion stems from possible confinement mechanisms. The Kugo-Ojima scenario~\cite{Kugo:gm} puts 
forward the idea that all colored objects are BRST charged and thus unphysical. One precondition for this mechanism is 
an unbroken global color charge. In the Landau gauge this condition can  be cast into \cite{Kugo:1995km}
\be 
\lim_{p^2 \to 0} p^2 D_G(p^2)\to\infty,\label{Kugo} 
\ee
where $D_G$ is the propagator of the Faddeev-Popov ghost. This scenario necessarily also implies the condition 
\pref{Oehme} for the gluon propagator.

In the Gribov-Zwanziger scenario~\cite{Zwanziger:2003cf}, entropy arguments are employed to show the dominance of 
field configurations close or on the Gribov horizon in field configuration space. This scenario predicts again 
condition \pref{Kugo}. For an infrared constant ghost-gluon vertex, which is supported by lattice calculations 
\cite{Cucchieri:2004sq} and semi-perturbative calculations \cite{Schleifenbaum:2004id}, condition \pref{Oehme} follows 
as well for the gluon propagator \cite{Zwanziger:2003cf}.

Also intuitively it is clear that a strongly divergent ghost propagator at zero momentum can mediate confinement. Such 
an infrared divergence relates to long-ranged spatial correlations. These are stronger than the ones induced by a 
Coulomb force since the divergence in momentum space is stronger than that of a massless particle. An infrared 
vanishing gluon propagator is also intuitively linked to confinement, as $p^2=0$ puts a non-interacting gluon 
on-shell. Thus the gluon does not propagate and is confined.

\section{Dyson-Schwinger equations}\label{sdse}

The DSEs~\cite{Dyson:1949ha,Alkofer:2000wg} form an infinite tower of coupled non-linear integral equations for the 
Green's functions of a given theory. Therefore, in general only a truncated set can be solved in practical 
calculations. In the following we aim at a closed set of equations for the pertinent 2-point functions. In Landau 
gauge and at finite temperature these are the ghost propagator
\be
D_G(p)=\frac{-G(p_0^2,\vec p^2)}{p^2}\label{ghostDressing}
\ee
and the gluon propagator \cite{Kapusta:tk}
\be
D_{\mu\nu}(p)=P_{T\mu\nu}(p)\frac{Z(p_0^2,\vec p^2)}{p^2}+P_{L\mu\nu}(p)\frac{H(p_0^2,\vec p^2)}{p^2} \; 
.\label{gluonDressing}
\ee
where $P_T$ and $P_L$ are projectors transverse and longitudinal w.r.t.\ the heat-bath. Eqs.\ \pref{ghostDressing} and 
\pref{gluonDressing} define the dimensionless dressing functions $G(p_0^2,\vec p^2)$, $Z(p_0^2,\vec p^2)$, and 
$H(p_0^2,\vec p^2)$.

The derivation of the Dyson-Schwinger equations is a straightforward, but tedious task. As in the previous 
investigations \cite{Maas:2004se,Gruter:2004bb}, we follow here ref.\ \cite{vonSmekal:1997is} and keep only the equations for the propagators. Furthermore we neglect 
one-particle-irreducible two-loop diagrams and assume a perturbative color structure. Furthermore, the ghost-gluon 
vertex is taken to be bare, in accordance with recent investigations \cite{Cucchieri:2004sq,Schleifenbaum:2004id}. The 
construction of the various three-gluon vertices will be discussed in section \ref{sstrunc}.

The equations are obtained from the vacuum equations (see e.g.\ ref.\ \cite{Alkofer:2000wg}) by application of the Matsubara 
formalism \cite{Kapusta:tk}. To obtain scalar equations for the (infinite) set of Matsubara frequencies of the 
dressing functions $Z$ and $H$ of the gluon propagator \pref{gluonDressing}, the gluon equation is contracted with the 
generalized projectors $P_{T\mu\nu}^\zeta$ and $P_{L\mu\nu}^\xi$, respectively, defined by\footnote{These are chosen 
differently from the ones in ref.\ \cite{Maas:2004se} to avoid an inconvenient kinematical singularity. In the limit 
of infinite temperature, the resulting equations for the soft modes are the same.}
\bea
P_{T\mu\nu}^\zeta&=&\zeta P_{T\mu\nu}+(\zeta-1)\left(\delta _{\mu \nu }-\delta _{\mu 0}\delta 
_{0\nu}\right)\label{gtproj},\\
P_{L\mu\nu}^\xi&=&\xi P_{L\mu\nu}+(\xi-1)\left(\delta_{\mu 0}\frac{p_0 p_\nu}{p^2}+\delta_{0\nu}\frac{p_\mu 
p_0}{p^2}\right)\label{glproj}.
\eea
\noindent The choice of the projectors was made so as to obtain a well-defined 3-dimensional limit. The parameters 
$\zeta$ and $\xi$ allow to vary the projection continuously in order to investigate the amount of gauge symmetry 
violations. The dependence on $\xi$ vanishes in the infinite-temperature limit \cite{Maas:2004se}.

\begin{figure*}
\epsfig{file=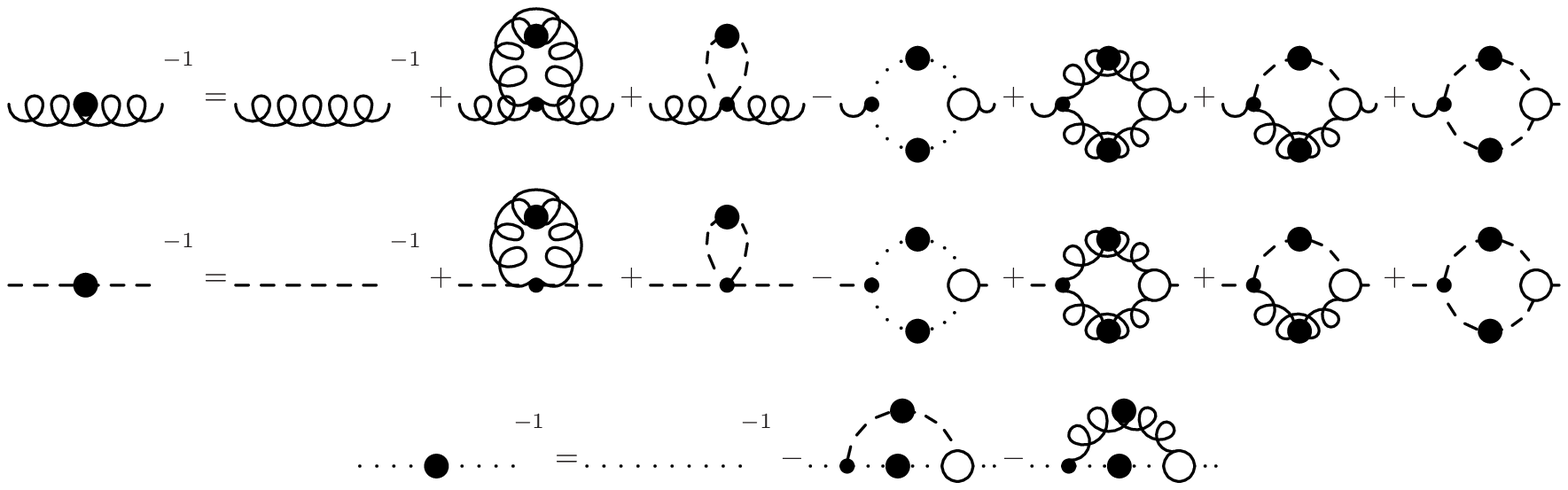,width=\linewidth}
\caption{The truncated Dyson-Schwinger equations at finite temperature. The dotted lines denote ghosts, the dashed 
lines longitudinal gluons and the wiggly lines transverse gluons. Lines with a full dot represent self-consistent 
propagators and small dots indicate bare vertices. The open circled vertices are full and must be constructed in a 
given truncation scheme. A bare ghost-gluon vertex and slightly modified bare gluon vertices have been used here. Note 
that for soft modes the ghost-longitudinal and the 3-point coupling of three longitudinal and of one longitudinal and 
two transverse gluons vanish.}\label{figsysft}
\end{figure*}

Employing the projectors \pref{gtproj} and \pref{glproj} yields the finite temperature DSEs as
\bea
\frac{1}{G(p)}&=&\widetilde{Z}_3\nonumber\\
&+&\frac{g^2TC_A}{(2\pi)^2}\sum_{n=-\infty}^{\infty}\int d\theta d\left| \vec q\right| 
\Big(A_T(p,q)G(q)Z(p-q)\nonumber\\
&&+A_L(p,q)G(q)H(p-q)\Big)\label{fulleqGft}\\
\frac{\xi}{H(p)}&=&\xi Z_{3L}+T^{HG}+T^{HH}\nonumber\\
&&+\frac{g^2TC_A}{(2\pi)^2}\sum_{n=-\infty}^{\infty}\int d\theta d\left| \vec q\right|\Big(P(p,q)G(q)G(p+q)\nonumber\\
&&+N_L(p,q)Z(q)Z(p+q)+N_1(p,q)H(q)Z(p+q)\nonumber\\
&&+N_2(p,q)H(p+q)Z(q)+N_T(p,q)H(q)H(p+q)\Big)\nonumber\\\label{fulleqHft}\\
\frac{1}{Z(p)}&=&Z_{3T}+T^{GH}+T^{GG}\nonumber\\
&&+\frac{g^2TC_A}{(2\pi)^2}\sum_{n=-\infty}^{\infty}\int d\theta d\left| \vec q\right|\Big(R(p,q)G(q)G(p+q)\nonumber\\
&&+M_L(p,q)H(q)H(p+q)+M_1(p,q)H(q)Z(p+q)\nonumber\\
&&+M_2(p,q)H(p+q)Z(q)+M_T(p,q)Z(q)Z(p+q)\Big)\nonumber\\
&&+\frac{p_0^2(\zeta-1)}{2p^2}\left(Z_{3L}-\frac{1}{H(p)}\right).\label{fulleqZft}
\eea
\noindent Here $\delta^a_d C_A=f^{dbc}f_{abc}=\delta^a_d N_{c}=\delta^a_d 3$ is the adjoint Casimir of the gauge group. The summation runs over all 
Matsubara frequencies $q_0=2\pi T n$. This (truncated) set of DSEs is graphically displayed in Fig.~\ref{figsysft}.

The $\zeta$ and $\xi$ dependence of the DSEs is acquired by using \pref{gtproj} and \pref{glproj}. At $\zeta=\xi=1$ 
the original form of the equations is recovered. For $p_0=0$, equation \pref{fulleqHft} is only superficially 
dependent on $\xi$. As all integral kernels are in this case proportional to $\xi$, the dependence can be divided out 
for $\xi\neq 0$. Then, only the implicit dependence through interactions with the hard modes remains. The latter 
effect vanishes in the infinite-temperature limit and \pref{fulleqHft} becomes the equation for the Higgs field of 
\cite{Maas:2004se}, independent of $\xi$.

The finite-temperature theory is renormalizable. Thus, explicit wave-function renormalization constants 
$\widetilde{Z}_3$, $Z_{3L}$, and $Z_{3T}$ have been introduced and will be discussed in section \ref{sstrunc}. 
Concerning the vertex renormalization,  $\widetilde{Z}_1=1$ has been employed for the ghost-gluon vertex 
\cite{Taylor:ff}. The three-gluon vertex renormalization constant $Z_1$ cannot be calculated in the present truncation, 
but it is finite and set to 1 \cite{Fischer:2002hn,Fischer:2003zc}. The kernels $A_T$, $A_L$, $R$, $M_T$, $M_1$, 
$M_2$, $M_L$, $P$, $N_T$, $N_1$, $N_2$, and $N_L$ contain trivial factor such as the integral measure. The integral 
kernels are quite lengthy, and will not be displayed here. They can be found in ref.\ \cite{Maas:2005rf}.

The tadpoles $T^{ij}$ are used to cancel spurious divergences in much the same way as in ref.\ \cite{Maas:2004se}, see 
section \ref{sstrunc}. However, also in the soft equations, they can possess finite parts at finite temperature. In 
case of the longitudinal equation, these can be absorbed into the mass renormalization discussed below. In the 
transverse equation, these are at $\zeta=3$ completely contained in the loop terms, and their continuation away from 
$\zeta=3$ is thus arbitrary. Also these contributions scale at best only as $1/p^2$ in the infrared and thus turn out 
to be irrelevant. Therefore, they are dropped, especially as no simple prescription as the one given later in 
Eq.~\pref{finite} can remove the related spurious divergences

Close inspection of the equations reveals further that the dressing functions can only depend on $\left|p_0\right|$ 
and $\left|\vec p\right|$. The corresponding symmetry, under $p_0\to -p_0$, is used to reduce the number of equations 
significantly.

Eqs.~\prefr{fulleqGft}{fulleqZft} depend on the coupling constant $g$ as the only parameter. In turn, $g$ depends on 
the renormalization scale $\mu$, which can be chosen arbitrarily at any fixed temperature $T$. It directly enters into 
the definition of the infinite-temperature limit, as the effective 3-dimensional coupling constant $g_3$ depends on 
$g$ \cite{Kajantie:1995dw}. In the simplest case, $g_3^2\sim g^2(\mu)T$. Comparing Eqs.~\prefr{fulleqGft}{fulleqZft} 
to those of the infinite-temperature limit \cite{Maas:2004se}, the constant of proportionality has to be 1 in this 
truncation scheme in order to obtain a smooth infinite-temperature limit. Hence, it only remains to choose the 
temperature dependence of $\mu$. As discussed in appendix \ref{appcc} a t'Hooft-like scaling \cite{'tHooft:1973jz} 
with $g^2T$ fixed is employed to obtain a smooth infinite-temperature limit.

\section{Infrared properties}\label{sir}

Asymptotic freedom turns out to be advantageous for the infrared as well. Just at the phase transition, the $n=1$ 
Matsubara frequency has already an effective `mass' $p_0=2\pi T_c\approx 1.7$ GeV. By the Appelquist-Carazzone theorem 
\cite{Appelquist:tg} the hard modes are suppressed by powers of $|\vec p|/p_0$ in the infrared. Thus, hard modes do 
not contribute significantly. Therefore, the infrared behavior of the soft modes is the same as in the 
infinite-temperature limit and given by power-laws,
\bea
G(p)&=& A_g p^{-2\kappa},\nonumber\\
Z(p)&=& A_z p^{-2t},\nonumber\\
H(p)&=& A_h p^{-2l}.\nonumber
\eea
\no The longitudinal function $H$ behaves mass-like with $l=1$. The exponents of the ghost and transverse gluon 
propagators are related as \cite{Maas:2004se,Zwanziger:2001kw}
\be
\kappa=-\frac{1}{2}\left(t+\frac{1}{2}\right).\nonumber
\ee
\no As in the infinite-temperature limit, always two solutions are found. One has $\kappa=1/2$ while the other has a 
weakly $\zeta$-dependent value, being $\kappa=0.39760$ at $\zeta=1$ \cite{Maas:2004se,Zwanziger:2001kw}. The later is 
the more likely solution, at least in the infinite-temperature limit \cite{Maas:2004se}. Hence the infrared exponents 
are independent of temperature and therefore at least gluons transverse w.r.t.\ the heat-bath are confined above the 
phase transition. Studying each equation in detail, it turns out that these quite general statements are implemented 
very differently in each equation.

In case of the hard modes, there is no purely soft contribution due to momentum conservation at the vertices. Thus 
they decouple and a self-consistent solution is that all hard-mode dressing functions are constant in the infrared.

In the case of the soft ghost equation \pref{fulleqGft}, all contributions become constant in the infrared. Thus they 
can be canceled by the (renormalized) tree-level value. Only the subleading behavior remains, as in the case of the 
3-dimensional  and 4-dimensional theory \cite{Maas:2004se,vonSmekal:1997is}. As the subleading contribution of the 
purely soft term dominates the hard terms, the same behavior as in the infinite-temperature limit emerges.

In case of the transverse equation \pref{fulleqZft}, the hard mode contributions give rise to mass-like $1/p^2$ terms
at best. These terms are subleading compared to the soft ghost-loop. Thus the infrared is dominated by the latter and 
the same infrared solution as in the infinite-temperature limit is found.

Finally, the longitudinal equation \pref{fulleqHft} is quite different from its equivalent in the 3-dimensional  or 
4-dimensional case. In the prior case, it was dominated by its tree-level mass, while in the latter it is identical to 
the transverse one. In the present case, the absence of pure soft interactions with ghosts, which could generate 
divergent contributions as in the transverse equation, leads to dominance of the hard modes\footnote{The pure soft 
interactions represented by the kernels $N_1$ and $N_2$ only generate a constant term in the infrared. Without a 
tree-level mass, the soft tadpoles generate a contribution which can be removed by renormalization.}. These provide 
mass-like contributions in the infrared. Inspecting e.g. $P$ at $\xi=1$
\be
P(0,q_0,\vec q,\vec p)=\frac{\left|\vec q\right|^2q_0^2\sin\theta}{\left|\vec p\right|^2q^2(q_0^2+(\vec p+\vec 
q)^2)},\nonumber
\ee
\noindent and using the fact that the hard dressing functions are constant in the infrared, directly leads to a 
mass-like behavior due to the explicit $1/|\vec p|^2$ factor and the finiteness of the remaining expression. Thus, 
the 3-dimensional mass of the Higgs sector is generated dynamically by the interaction with the hard modes.

In conclusion, the infrared exponents of all soft mode dressing functions do not depend on the temperature. Only the 
infrared coefficients change.

These observations imply that, by using a finite number of Matsubara frequencies, it will not be possible to obtain 
the infrared 4-dimensional vacuum solutions. This would only be possible, if the infrared limit of the high-temperature and the low-temperature phase would have been both encoded in the soft interactions alone or, as the electric screening mass, would already be induced by a single hard mode. However, any finite number of modes will not be able to generate a divergence stronger than mass-like in the longitudinal equation, equivalent to the soft ghost contribution in the transverse equation. Hence, the 4-dimensional behavior can be established only by infinite summation. In this case, the generated mass has to diverge 
even after renormalization at $p=0$ to obtain the vacuum solution, providing over-screening instead of screening. The 
possibility that the ghost-gluon vertex changes due to temperature resulting in a coupling of soft ghost modes to soft 
longitudinal gluons at low temperature but not at high temperature is unlikely, since a bare ghost-gluon vertex is 
already sufficient at low temperatures and in the vacuum \cite{Gruter:2004bb,vonSmekal:1997is}.

\section{Truncation}\label{sstrunc}

Concerning the truncation chosen, all arguments proven to be successful in the vacuum and the infinite-temperature 
limit \cite{Maas:2004se,vonSmekal:1997is,Maas:2005rf} also apply here. This includes the requirement that, due to the 
Gribov condition, the dressing functions $G$, $Z$, and $H$ have to be positive semi-definite. Hence the discussion 
here will only cover the additional aspects due to finite temperature. The main additional truncation is to include 
only the finite set $[-N+1,N-1]$ of Matsubara frequencies.

For large but still finite $T$  the 3-dimensional limit is to be made explicit. Thus, all the problems encountered in 
the 3-dimensional theory \cite{Maas:2004se} persist, including the necessity of a modified soft 3-gluon vertex. Hence, 
all modifications\footnote{Recent investigations indicate that the three-gluon vertex may have an infrared divergence 
\cite{Alkofer:2004it}. As pointed out also in ref.\ \cite{Alkofer:2004it}, this does not affect the infrared regime, 
as this range is dominated by the ghost loop. This is likely also the case at finite temperature in the transverse 
equation. Due to the chromoelectric screening it is conceivable that this behavior does not persist in the 
longitudinal equation at high temperatures.} of the pure soft terms will be left as in the case of \cite{Maas:2004se}, 
except for the tadpoles in the Higgs equation \pref{fulleqHft}. Due to the dynamically generated mass, it will be 
necessary to alter this behavior. As no further information on the interaction vertices involving hard modes is 
available, the corresponding vertex functions will be assumed bare. Due to the large mass of the hard modes, such a 
tree-level ansatz should be justified.

The hard mode contributions induce additional spurious divergences. As they contribute to the two-point functions 
mostly at mid-momenta, when viewed from the 4-dimensional perspective using $p^2=p_0^2+\vec p^2$, it is expected that 
in the chosen truncation, they are much harder to compensate. This is indeed the case. The subtraction prescriptions 
are listed in appendix \ref{appsub}. The technically quite involved construction is described in detail in ref.\ 
\cite{Maas:2005rf}. 

After removal of the spurious quadratic divergences, the resulting equations are still logarithmically divergent for 
$N\to\infty$. The integrals themselves are convergent, but the Matsubara sum is not, as the terms scale like $1/q_0$. 
Hence the sum diverges for an infinite number of Matsubara frequencies. This is the 4d logarithmic divergence and thus 
the usual one of Yang-Mills theory \cite{Bohm:yx}, which has to be renormalized appropriately. To this end, three 
regions of momentum are to be distinguished:

At $p\le 2\pi T$, all terms in the Matsubara sums behave essentially the same due to the Appelquist-Carrazone 
decoupling theorem. It is valid as the hard modes behave essentially tree-level-like. As the contributions of the hard 
modes behave as $1/q_0$, the final result will depend on the number of Matsubara frequencies included. Thus, it is 
necessary to renormalize in order to be independent of the cutoff, which is in this case imposed by the number of 
Matsubara frequencies included. The renormalization procedure will be implemented below. By this approach, the results 
can be made quite reliable in this regime.

At $2\pi T\le p\le 2\pi T (N-1)$, more and more Matsubara terms depart from their $1/q_0$ behavior to a $1/p$ 
behavior. As the external momentum $p$ becomes large compared to the effective mass $q_0$ of a hard mode, the mode 
becomes dynamical and behaves like a massive 3-dimensional particle. The results are still quite reliable in this 
region, since also for $N\to\infty$, at any finite momentum $p$ only a finite number of Matsubara frequencies are 
dynamical. It becomes less and less reliable when approaching the upper limit $2\pi T(N-1)$.

At $p\ge 2\pi T (N-1)$, the situation changes drastically. Opposite to the case $N=\infty$, all Matsubara modes are 
dynamical and their contributions will scale as $1/p$. Thus the number of Matsubara modes will now enter linearly 
instead of logarithmically. This is an artifact of cutting off the Matsubara sum. In this region the results are not 
reliable. However, as the sum is still finite and suppressed by $1/p$, the contribution is subleading with respect to 
the tree-level term and thus the system of equations can still be closed consistently\footnote{A further argument is 
that otherwise the corresponding finite 3-dimensional theory would be ill-defined which is not the case.}.

By renormalization, these artifacts can be reduced, if not completely removed at sufficiently small momenta. In that 
sense the finite Matsubara sum approximation is a small-(3-)momentum approximation. However, there are a few 
subtleties involved concerning the renormalization of a truncated Matsubara sum.

In the vacuum case \cite{Alkofer:2000wg,Fischer:2002hn}, the DSEs can be renormalized using a momentum subtraction 
scheme (MOM). This approach fails if the large momentum asymptotic value of the dressing functions is a constant 
different from 0. In the case of a truncated Matsubara sum, the ultraviolet behavior is that of a massive 
3-dimensional theory. Thus the self-energy contributions vanish as $1/p$ in the ultraviolet \cite{Maas:2004se}. 
Therefore the dressing functions $F=G$, $Z$, $H$ are dominated by the tree-level term
\be
\lim_{\left|\vec p\right|\to\infty}F(p_0,\left|\vec p\right|)\to\frac{1}{Z_3}.\nonumber
\ee
\noindent Here $Z_3$ is generically the wave function renormalization of $F$. For a finite number of Matsubara frequencies $Z_3$ is finite. Therefore the renormalization will here be 
performed by explicit counter-terms. This is discussed below.

A second point is the mass and mass renormalization necessary for the soft longitudinal mode. The soft mode with 
frequency $p_0=0$ of the $A_0$ component of the gauge field transforms homogeneously instead of inhomogeneously under 
gauge transformations. Therefore gauge symmetry permits to add a term
\be
\delta m^2A_0^2(0,\vec p)\label{a0mass}
\ee
\noindent to the Lagrangian. In the vacuum such a term is forbidden by manifest Lorentz invariance. At finite 
temperature in the Matsubara formalism, this is no longer the case, and such a term could in principle be present. 
This term replaces the $p_0^2A_0^2$ term of the hard modes, which stems from the $A_0\pd_0^2 A_0$ term in the 
Lagrangian for $p_0\neq 0$.

Concerning the counter-terms, the wave-function renormalization is performed by adding the counter-term
\be
\delta Z_3(A_\mu^a\pdm\pdn A_\nu^a-A_\mu^a\pd^2 A_\mu^a)\nonumber
\ee
\noindent to the Lagrangian. In the case $p_0=0$, the first term is not present for the longitudinal mode $A_0$. Its 
place is taken by \pref{a0mass}. This implies a relation between the wave-function counter-term $\delta Z_3$ and the 
mass counter-term $\delta m^2$, which cannot be exploited here due to the truncation of the Matsubara sum. Therefore 
an independent renormalization of the wavefunction and the mass of the soft longitudinal mode is necessary and will be 
performed.

A last point concerns the implications for the counter-terms due to the truncation of the Matsubara sum. As the 
divergence structure must be the same as in the vacuum, the counter-terms must be the same for all frequencies. This 
is no longer the case when the Matsubara sum is truncated, as can be seen directly by counting. In the sum for the 
soft mode, $p_0=0$, contributions from $2N-1$ Matsubara modes are present for each loop. For the hard mode with 
$p_0=2\pi T(N-1)$, only $N$ contributions are present, as $\left|p_0+q_0\right|\le 2\pi (N-1)T$. Thus different 
numbers of modes contribute and the counter-terms cannot be the same. This is always the case as long as $N<\infty$. 
To surpass the problem in a constructive manner, each mode will be renormalized independently.

Therefore, a counter-term Lagrangian is added, given by
\bea
\delta{\cal L}&=&\delta m^2 A_0^a(0)^2+\sum_{q_0} \Big(\delta Z_{3T}(q_0) A^a_\mu(q_0) \Delta_{T\mu\nu}(q_0) 
A^a_\nu(q_0)\nonumber\\
&+&\delta Z_{3L}(q_0) A^a_\mu(q_0) \Delta_{L\mu\nu}(q_0) A^a_\nu(q_0)\nonumber\\
&+&\delta\widetilde{Z}_3(q_0) \bar c^a(q_0)\pd^2 c^a(q_0)\Big)\label{ctl}
\eea
\noindent where $A^a_\mu(q_0)$ are the modes of the gluon field and $\bar c^a(q_0)$ and $c^a(q_0)$ are the modes of the 
ghost and anti-ghost field, respectively. $\Delta_{T/L\mu\nu}$ are the appropriate tensor structures of derivatives.

\begin{figure}
\epsfig{file=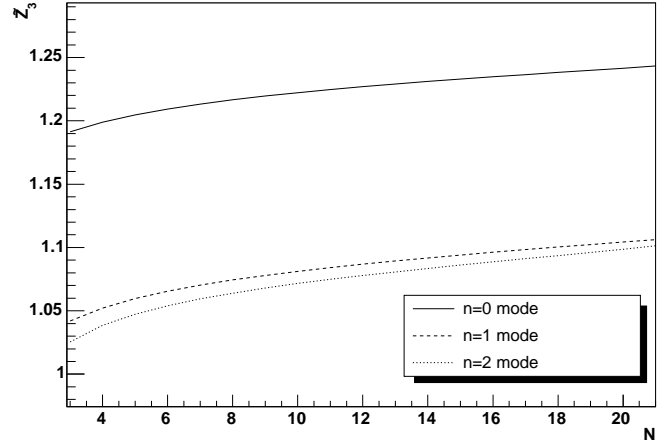,width=\linewidth}
\caption{Dependence of $\widetilde{Z}_3$ on the number of Matsubara frequencies $N$ for the three lowest frequencies. 
Shown are the values for the $\kappa=1/2$ solution at $T=1.5$ GeV. The solution for $\kappa\approx 0.4$ is very similar, even on 
a quantitative level.}\label{figz3}
\end{figure}

This shortcoming is a consequence of the truncation of the Matsubara sum, yielding incorrect ultraviolet properties: 
For large momenta the system under consideration is equivalent to a 3-dimensional theory of $2N-1$ particles per 4d 
particle species. In such a theory all fields can be renormalized independently. When the limit $N\to\infty$ is 
performed, the wavefunction renormalizations will again coincide, as is shown\footnote{For gluons, the value of the 
wavefunction renormalization, in the computationally accessible range of $N$-values, is dominated by $N$-dependent 
effects at mid-momenta. Hence, the logarithmic running is not easily discernible.} for $\widetilde{Z}_3$ in Fig. 
\ref{figz3}. However, this process is logarithmically slow.

In all equations, but the one for $H(0,\vec p)$, using the counter-terms of \pref{ctl} amounts to replacing the 
tree-level term $1$ by $1+\delta Z_3=Z_3$. In this way multiplicative renormalization is obtained explicitly. In the 
equation for $H(0,\vec p)$, the tree-level term is replaced\footnote{Note that when using the projector \pref{glproj}, 
also $\delta m^2$ is multiplied by $\xi$.} by $1+\delta Z_{3L}+\delta m^2/p^2$. Thus, it also generates a 
mass-renormalization.

The last ingredient is the renormalization prescription, which will be taken as
\be
F(s)=1,\nonumber
\ee
\noindent where $s$ is the subtraction point. Here $s=T$ is chosen, see \cite{Maas:2005rf} for an alternative. This 
also ensures $G(s)^2Z(s)=1$ as is required in the 4-dimensional theory \cite{vonSmekal:1997is}.

In case of $H(0,\vec p^2)$, two prescriptions are necessary. The first is\footnote{For numerical reasons, it is 
actually performed at $2\delta_i$, where $\delta_i$ is the numerical IR-cutoff of the integration. This causes only a 
marginal difference.}
\be
\lim_{|\vec p|\to 0}|\vec p|^{-2}H(0,\vec p^2)=\frac{1}{m_{3d}^2}=\frac{1}{r^2g^4T^2+g^2TC_A\frac{rg^2T}{4\pi}}\nonumber
\ee
\noindent where $m_{3d}$ is the tadpole-improved mass of the 3-dimensional theory \cite{Maas:2004se}. It depends on 
$r=m_h/g_3^2$, which is again taken to be the same value as in ref.\ \cite{Maas:2004se}. For fixed $g_3$, this mass is 
independent of temperature. In addition,
\be
\frac{1}{H(0,s)}=1+\frac{m_{3d}}{s^2}\nonumber
\ee
\noindent is required to fix the wavefunction renormalization.

It should be noted that, by renormalizing at $T$, it is guaranteed that the correct 3-dimensional limit is obtained. 
This prescription requires that at $T\to\infty$ all dressing functions approach 1 at infinity \cite{Maas:2004se}.

Therefore, the explicit implementation of the renormalization prescription for the DSE of a dressing function $F$ with 
self-energy contributions $I$
\be
\frac{1}{F(p)}=1+I(p)\nonumber
\ee
\noindent is then
\bea
\frac{1}{F(p)}&=&1+\delta Z_3+I(p)\nonumber\\
\delta Z_3&=&-I(s)\nonumber
\eea
\noindent and in the case of $H(0,\vec p^2)$
\bea
\frac{1}{H(0,\vec p^2)}&=&1+\delta Z_{3L}+\frac{\delta m^2}{p^2}+I(p)\nonumber\\
\delta m^2&=&m_r^2-\lim_{p\to 0}p^2I(p)\nonumber\\
\delta Z_{3L}&=&-I(s)+\frac{\lim_{p\to 0}p^2I(p)}{s^2},\nonumber
\eea
\noindent where $m_r=m_{3d}$ is the renormalized mass.

This renormalization prescription amounts in total to a modified momentum subtraction scheme.

\section{Numerical results}\label{snum}

\subsection{Propagators}

Here the full system of Eqs. \prefr{fulleqGft}{fulleqZft} at all momenta will be treated\footnote{A separate discussion of a further truncation, the ghost-loop only truncation, can be found in ref.\ 
\cite{Maas:2005rf}. This truncation includes only tree-level and loop diagrams with at least one ghost-line.
These results support that it is indeed the ghost sector which drives the infrared behavior of Yang-Mills theory, also at finite 
temperature, in accordance with the Gribov-Zwanziger scenario.}. The numerical method employed is 
discussed in ref.\ \cite{Maas:2005rf,Maas:2005xh}.

\begin{figure*}
\epsfig{file=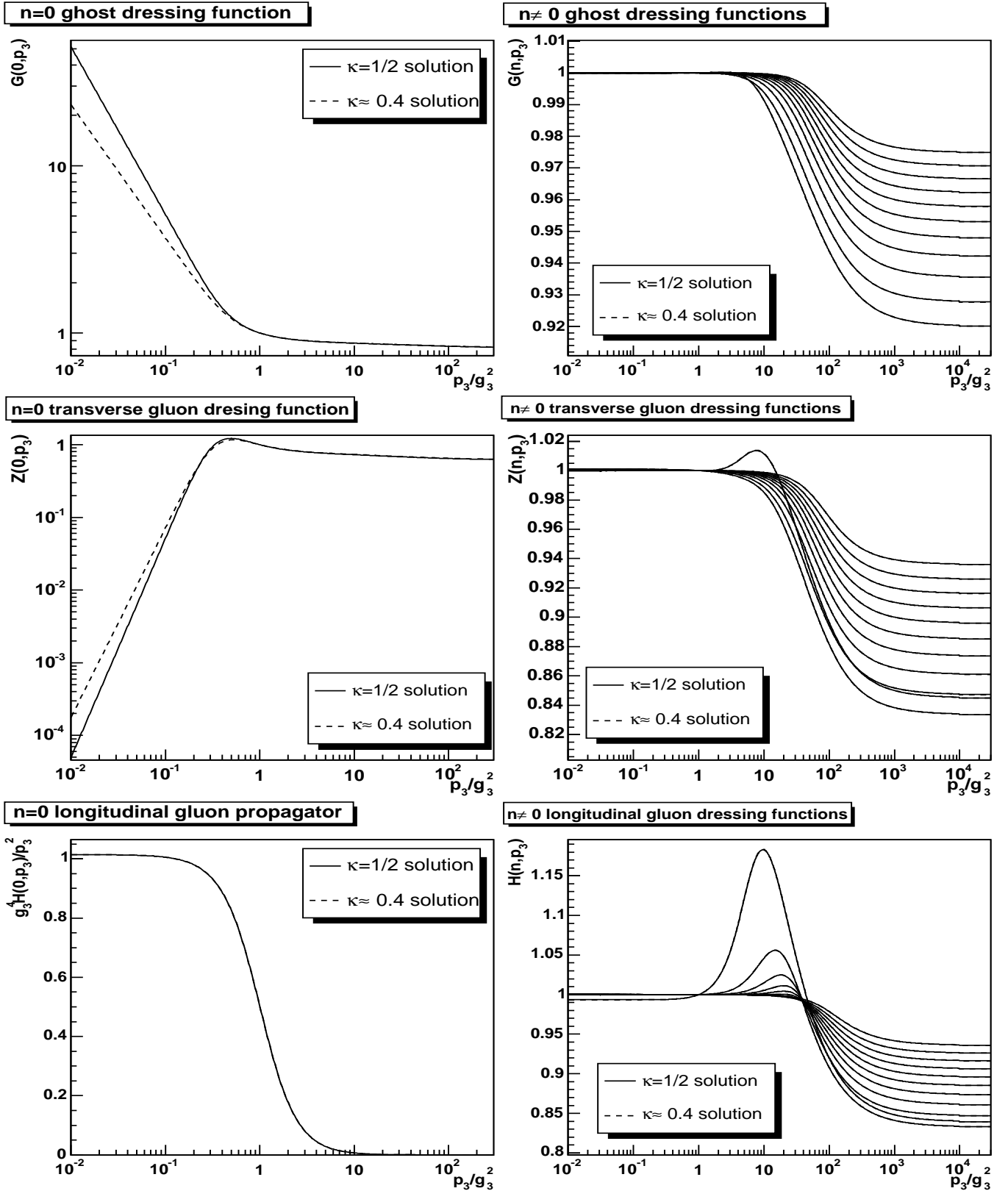,height=1.2\linewidth,width=\linewidth}
\caption{From top to bottom, the left panels show the dressing functions for the soft modes of the ghost and 
transverse gluon and the propagator of the longitudinal gluon. The right panels display the hard-mode dressing 
functions $n=1$ to $n=11$. Higher $n$ correspond to lesser deviation from unity except for $n=1$, see text. The solid 
lines are the $\kappa=1/2$ solution and the dashed lines the $\kappa\approx 0.4$ solution. Both are taken at $T=1.5$ GeV, 
$\zeta=\xi=1$, and $N=12$. Note the different momentum scale in the left and the right panels.}
\label{figftf}
\end{figure*}

For $N=12$ independent Matsubara frequencies, the result is shown in Fig. \ref{figftf} for both sets of infrared 
exponents, $\kappa=1/2$ and $\kappa=0.39760$. There are several observations. First of all, in the infrared the soft 
modes are nearly unaffected by the presence of the hard modes. The latter show a significant modification, compared to 
tree-level, although still only of the order of 30\%. The plateau reached in the ultraviolet is an artifact of the 
truncation, yielding a finite wave-function renormalization. The hard-mode dressing functions exhibit some structure. 
There are maxima in all dressing functions at mid-momenta. This is most pronounced in the case of the dressing 
functions of the longitudinal gluon. These structures do not translate into a corresponding structure in the 
propagators, which are monotonically decreasing from a constant of order $1/p_0^2$ in the infrared to 0 in the 
ultraviolet. For the hard-mode dressing functions it is also nearly irrelevant to which of the two soft infrared 
solution they belong.

In general, the hard modes deviate less from unity with increasing $n$. The only exception seems to be the ultraviolet 
behavior of the $n=1$ gluon modes. This is, however, an artifact of the truncation. If the Matsubara sum would not be 
truncated, all dressing functions would go to zero for sufficiently large momenta. On the other hand, the finite value 
attained due to the truncation can be influenced by the requirement to fit continuously to the mid-momentum behavior. 
Hence, due to the peaks at mid-momenta in the gluon dressing functions, the ultraviolet plateau is increased for the 
$n=1$ dressing functions, leading to the level reordering. If the peaks would be larger, this 
would also affect further modes. In the limit of $N\to\infty$, the peaks reach a finite 
maximum and therefore permit the plateaus all going to zero, thus restoring the correct behavior.

\begin{figure*}
\epsfig{file=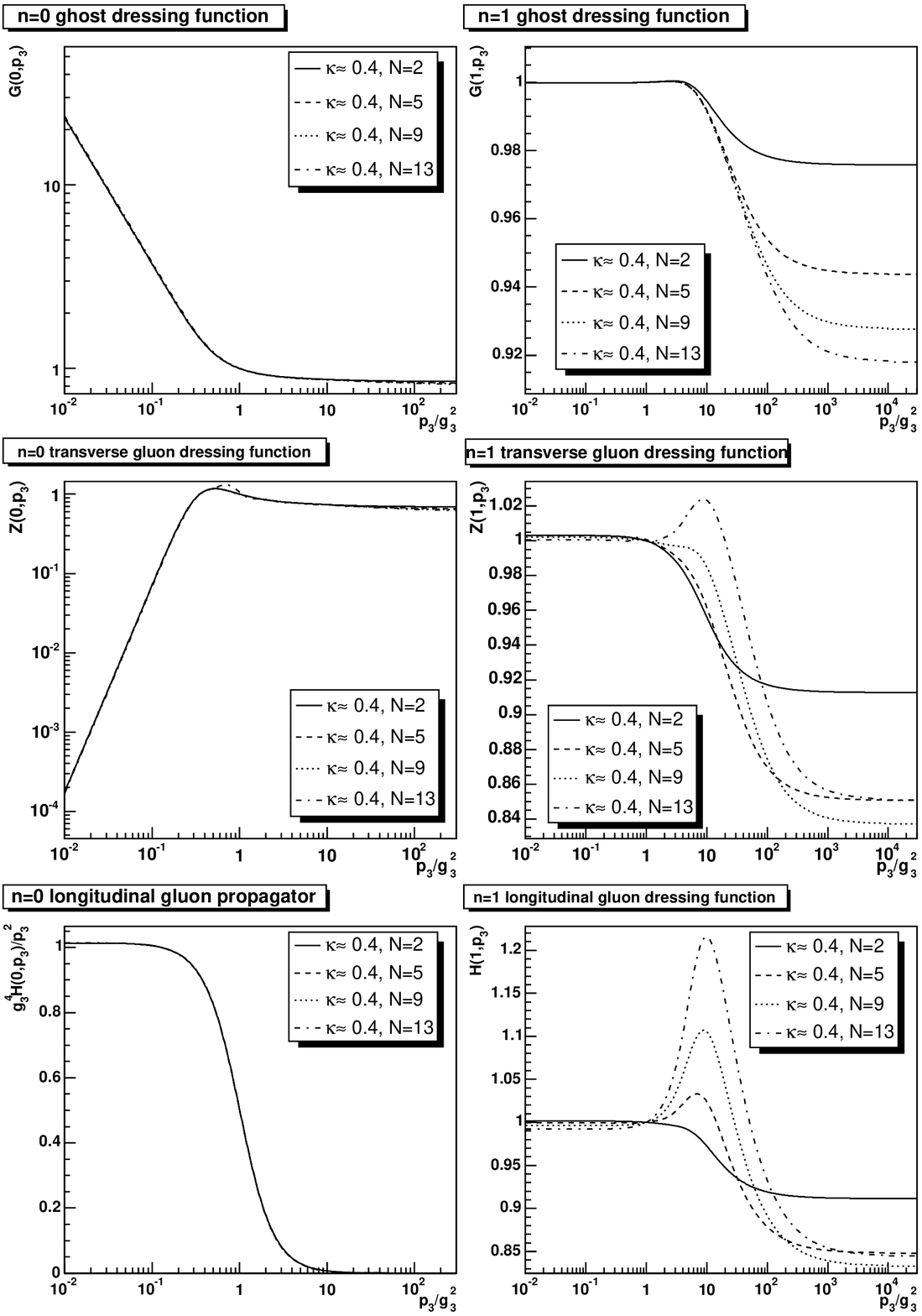,height=1.2\linewidth,width=\linewidth}
\caption{The dependence of the solutions on $N$ at $T=1.5$ GeV and $\zeta=\xi=1$. For the soft modes from top to 
bottom the left panels show  the dressing functions of the ghost and transverse gluon and the propagator of the 
longitudinal gluon. The right panels give the dressing functions for the $n=1$ hard mode. Only the $\kappa\approx 0.4$ 
solution is shown. The $\kappa=1/2$ solution is similar and can be found in ref.\ \cite{Maas:2005rf}. The solid lines 
denote $N=2$, dashed are $N=5$, dotted $N=9$ and dashed-dotted $N=13$.}
\label{figftfm}
\end{figure*}

Figure \ref{figftfm} displays the dependence of the full solutions on the number of Matsubara frequencies. While the 
soft-mode dressing functions are nearly unaffected, apart from the value of the renormalization constants, the effect 
on the hard mode dressing functions is significant. The ghost is quite insensitive, except for its wave-function 
renormalization. This is not the case for the gluons. The peaks at mid-momentum are sensitive to the number of 
Matsubara frequencies included. The effect is largest for the longitudinal gluon dressing functions. From the 
available number of Matsubara frequencies, it is hard to estimate whether the peak grows to a finite value for an 
infinite number of Matsubara frequencies or not. It therefore cannot be excluded that the hard longitudinal gluon 
dressing functions violate the Gribov condition, once sufficiently many Matsubara frequencies are included. This would 
be very similar to the case of the transverse gluon dressing function in the infinite-temperature limit 
\cite{Maas:2004se} and therefore would necessitate a similar vertex construction for the 
longitudinal-gluon-transverse-gluon vertex as for the soft transverse gluon vertex. If such a vertex would be 
necessary, the result would be a finite peak, very similar to the present situation. Thus, the results would be even 
quantitatively quite similar.

The disappearance of the peaks in the gluon dressing functions at $N=2$ can be directly related to the vanishing of 
hard-mode couplings due to the restriction $(p_0+q_0)/(2\pi T)<2$. Only hard-soft-mode couplings contribute. Thus, the 
peak is generated solely due to pure hard-hard interactions.

In general, the $n>1$ modes essentially follow the behavior of the $n=1$ mode, albeit much closer at tree-level.

\begin{figure*}
\epsfig{file=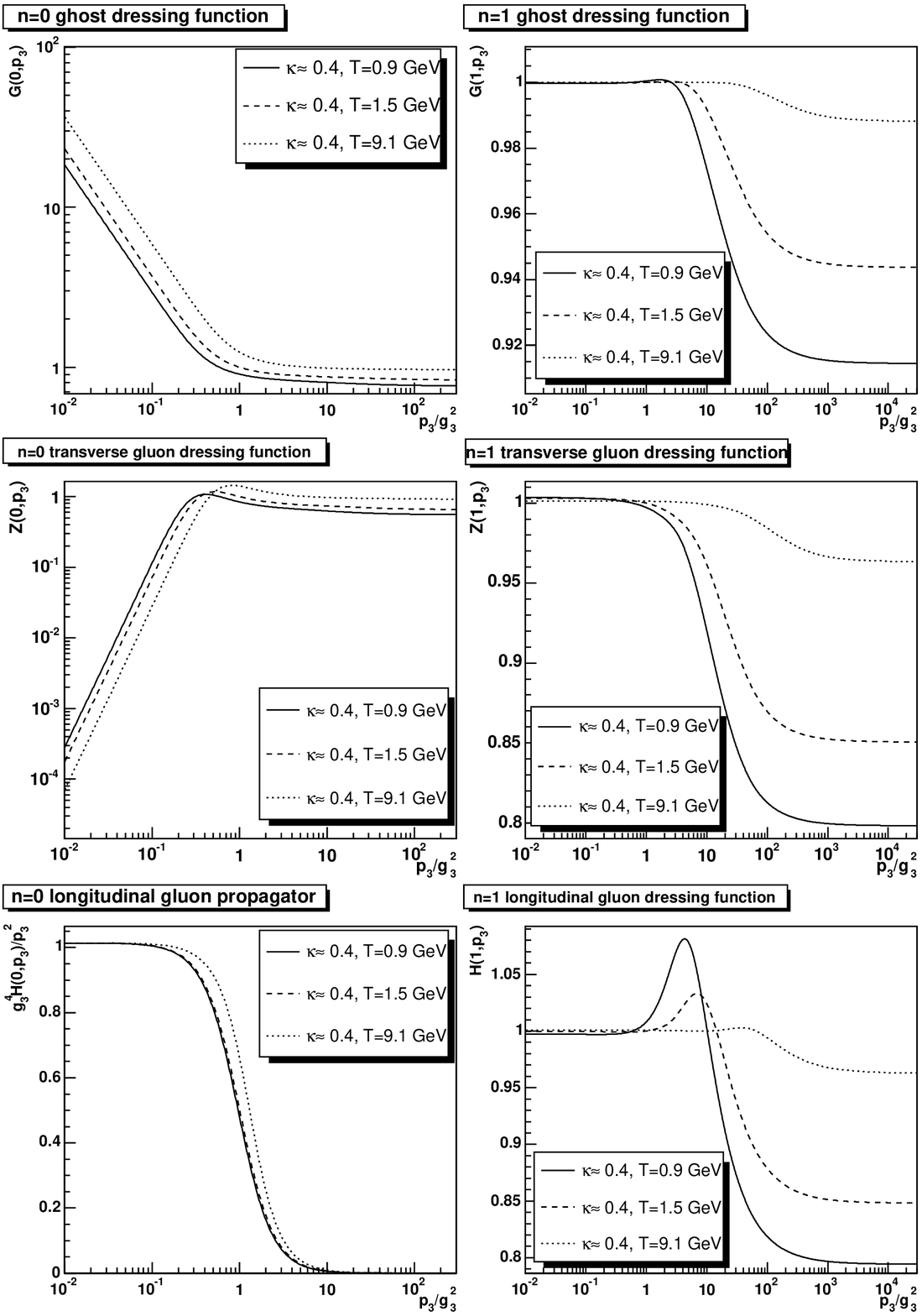,height=1.2\linewidth,width=\linewidth}
\caption{The dependence of the solutions on $T$ at $N=5$ and $\zeta=\xi=1$. The left panel shows the soft modes, where 
apart from the longitudinal gluon the dressing functions are shown. In the latter case the propagator is shown. The 
right panels show the dressing functions for the $n=1$ hard mode. Only the $\kappa\approx 0.4$ solution is shown, the 
$\kappa=1/2$ solution is similar and can be found in ref.\ \cite{Maas:2005rf}. Solid lines denote $T=0.9$ GeV, dashed lines 
$T=1.5$ GeV, and dotted lines $T=9.1$ GeV.}
\label{figftfg}
\end{figure*}

The dependence on temperature is shown in Fig. \ref{figftfg}. The infrared coefficients for the soft modes of the 
ghost and transverse gluon dressing functions are affected, while the dressing function of the longitudinal gluon only 
changes slightly at mid-momenta. Therefore, the dressing function of the longitudinal gluon is dominated by its 
renormalized mass also when changing the temperature. As expected, the hard-mode dressing functions become more and 
more tree-level like with increasing temperature. The peaks in the hard mode dressing functions shift to higher 
momenta, owing to the renormalization condition, and become smaller due to the increase of the effective mass $p_0$ of 
the hard modes. At infinite temperature, they smoothly merge into the corresponding 3-dimensional solutions. At low 
temperatures, the opposite effect is observed. In general, the hard-mode dressing functions become more dynamical as 
their effective mass decreases, albeit quite slowly. The insensitivity on the infrared solution of the soft sector is 
not changed when reducing the temperature, and it may need a significantly lower temperature to induce a change.

Employing t'Hooft scaling, the lowest temperatures that can be achieved with a numerically stable solution are of the 
order of $750$ MeV. Using a temperature-independent renormalization scale and subtraction point \cite{Maas:2005rf}
instead results merely in a shift in the onset of nonperturbative effects to larger momenta with some deformations, 
necessary to obey the renormalization conditions. In this case, it was also possible to cool down to $T=140$ 
MeV, indicating the possibility of super-cooling.

Varying the projection parameter $\zeta$ leads to similar variations for the soft-mode dressing functions as in the 
3-dimensional case \cite{Maas:2004se}. The largest effect is seen in the infrared. As the hard-mode dressing functions 
are insensitive to the infrared behavior of the soft modes due to their effective mass, there are only weak variations 
of them with $\zeta$. Only at small $\zeta$ an additional structure appears in the transverse gluon dressing function. 
This is likely due to the cross-term in Eq. \pref{fulleqZft}. The effect of varying $\zeta$ on the longitudinal sector 
is negligible, as in the 3-dimensional case. Correspondingly, the effect of varying $\xi$ is only significant for the 
longitudinal sector. The sensitivity is much less pronounced then in the case of varying $\zeta$. Especially, the soft 
longitudinal dressing function is nearly unaffected by variation of $\xi$. The reason is that the soft equation does 
not depend explicitly on $\xi$, as it can be divided out of the equation for $\xi\neq 0$. Therefore, any $\xi$ 
dependence enters only indirectly by the weak dependence of the hard modes on $\xi$. Only in the case of $\xi\approx 
0$ significant effects are found, owing to the pathology of the $\xi=0$ case. Details of the dependence on $\zeta$ and 
$\xi$ can be found in ref.\ \cite{Maas:2005rf}. The dependence on the three-gluon vertex construction is qualitatively 
not different from the infinite-temperature limit for the soft modes, and negligible for the hard modes 
\cite{Maas:2005rf}.

At this point, a comparison to a different approach to obtain the high-temperature gauge propagators can be made. The 
usual continuum method is the semi-per\-turbative hard thermal loop (HTL) approach \cite{Blaizot:2001nr}. It is based on 
resumming the hard mode contributions in self-energy diagrams. In the transverse infrared sector it is plagued by 
severe problems, due to its perturbative nature. The final result is a transverse gluon propagator with a particle 
like pole at $p=0$. Thus the gluon exponent $t$ would vanish. This is in sharp contrast to the results found here and, as 
discussed in ref.\ \cite{Maas:2004se}, such an infrared behavior is not very likely. Also the lattice results in the 
infinite-temperature limit \cite{Cucchieri:2001tw} and at temperatures somewhat above the critical 
\cite{Nakamura:2003pu} support the results found here. On the other hand, concerning the soft longitudinal mode and 
the hard modes, HTLs and the ansatz presented here find qualitatively similar results on the level of the propagators.

\subsection{Thermodynamic potential}\label{ssubtd}

Concerning the contribution from the hard modes to the thermodynamic potential, it is problematic to use the approach 
of Luttinger-Ward / Cornwall-Jackiw-Tomboulis (LW/CJT) \cite{Luttinger:1960ua}, as done previously \cite{Maas:2004se,Gruter:2004bb}. In the present truncation scheme the 2PI contribution in the 
LW/CJT-action are neglected, and thus only the interaction part is considered. This latter part vanishes identically for a free system, i.e.\ for a system 
containing only tree-level dressing functions. This free contribution has thus to be added explicitly. Therefore, the 
hard-mode contribution, being essentially tree-level, can be calculated up to perturbative corrections to be the same 
as that of a non-interacting system of free gluons. The LW/CJT-expression does not capture the tree-level contribution 
of the soft modes, and it thus can be added here. This yields a thermodynamic potential of a gas of massless gluons 
with small corrections due to the explicit soft-mode contributions and the residual interactions of the hard modes. 
For $N_c=3$ it is given by \cite{Karsch:2003jg} 
\be
\lim_{T\to\infty}\frac{\Omega}{T^4}=-\frac{16\pi^2}{90}.\label{sblimit}
\ee
\noindent Using t'Hooft-scaling for the interaction strength, the corrections due to the soft contributions obtained 
by the LW/CJT action vanish most likely as $1/T$ \cite{Maas:2004se,Maas:2005rf}. The only remaining contribution in 
the infinite-temperature limit is hence the Stefan-Boltzmann-contribution \pref{sblimit}, in agreement with results 
from lattice calculations \cite{Karsch:2003jg}. Therefore, the thermodynamic properties of the high-temperature limit 
are governed by the hard modes. However, as the calculation of the thermodynamic potential is problematic in the 
current approach \cite{Maas:2004se,Maas:2005rf}, this can only be taken as an indication for such a behavior.

At temperatures of the order of the phase transition, the contribution of the soft modes to the thermodynamic 
potential is probably non-negligible. Furthermore, as discussed in ref. \cite{Zwanziger:2004np}, these contributions 
can be highly relevant to the pressure $p$ and lead to a non-vanishing value for the trace of the stress-energy 
tensor, $\epsilon-3p$, where $\epsilon$ is the energy density. The vanishing of the trace would be expected for an 
ideal gas. Such an effect cannot be provided by perturbation theory \cite{Zwanziger:2004np}, but has been observed in 
lattice calculations \cite{Karsch:2003jg}.

\subsection{Schwinger functions}\label{ssschwinger}

\begin{figure*}
\epsfig{file=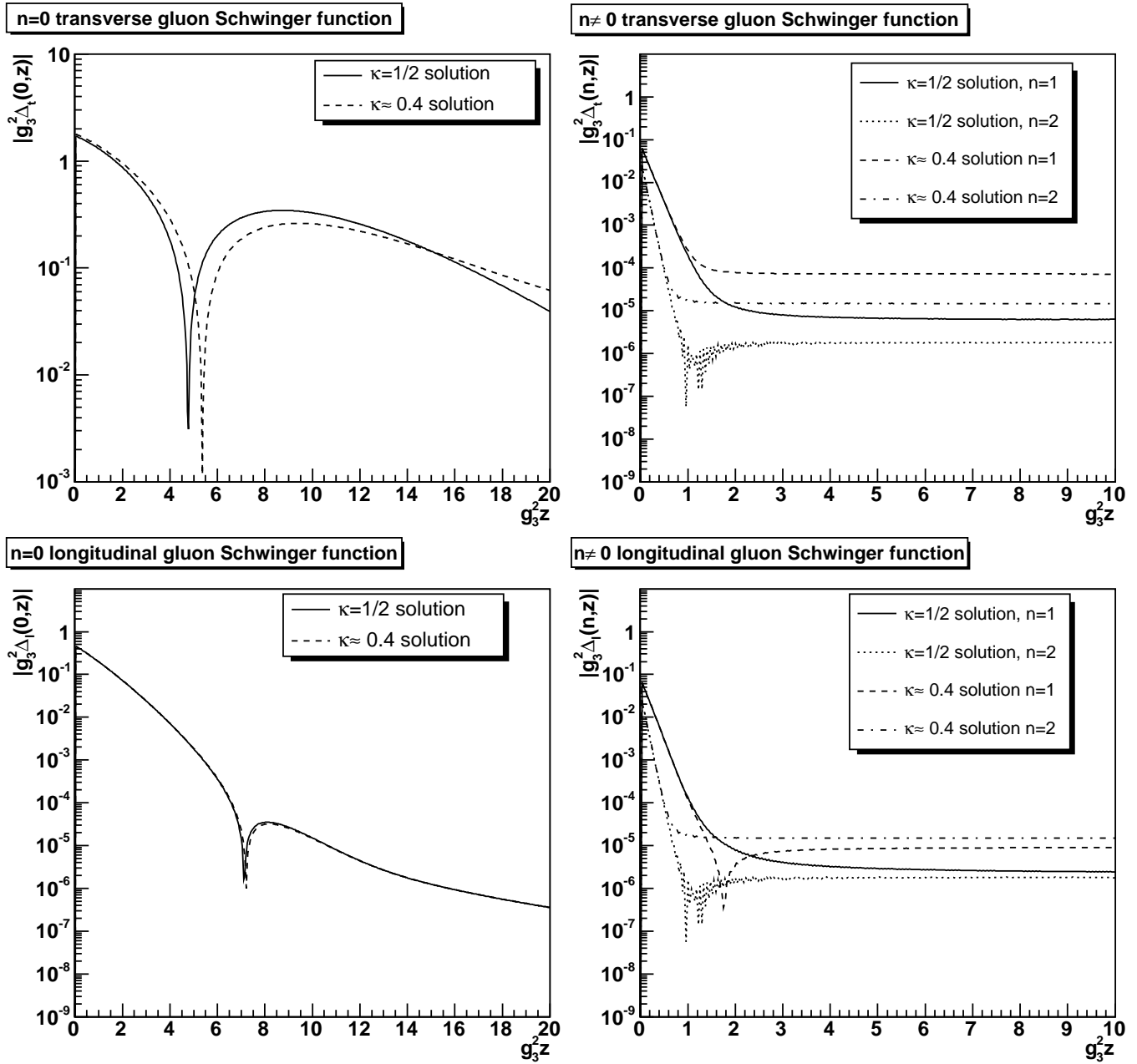,width=\linewidth}
\caption{The Schwinger function for the gluon propagator. The left panels show the results for the soft modes and the 
right panels for the hard modes with $n=1$ and $n=2$. The top panels give the results for the transverse propagator 
and the bottom panels for the longitudinal propagator. The temperature is $T=1.5$ GeV, $\zeta=\xi=1$, and $N=5$. The 
result has been obtained with roughly half a million Fourier frequencies \cite{Maas:2005rf}.}
\label{figschwinger}
\end{figure*}

As stated in section \ref{saspects}, an unambiguous sufficient (although not necessary) signal for the confinement of 
a particle is a violation of positivity.  This can be investigated using the Schwinger function, defined as
\be
\Delta(z)=\frac{1}{\pi}\int_0^\infty dp_0\cos(zp_3)\frac{F(p_3)}{p_3^2},\nonumber
\ee
\no where $F$ is either $Z$ or $H$. In the infinite-temperature limit a clear signal for positivity violation has been 
found for the transverse gluon \cite{Maas:2004se}, in agreement with lattice results \cite{Cucchieri:2001tw,Cucchieri:2000cy,Cucchieri:2004mf}. However, for the longitudinal gluon no unambiguous signal could be found. It is therefore interesting to repeat the study at lower temperatures. The results are shown in Fig.  
\ref{figschwinger}.

The soft transverse function does show a clear signal of positivity violation. The hard modes do not show a clear sign of oscillation within the available precision. Compared to the 
high-temperature case, a much clearer signal for positivity violation is found in the soft longitudinal propagator. 
Furthermore, the longitudinal Schwinger function is much more similar to the transverse one. Therefore, the 
contribution from the hard modes seem to be negative, shadowing the oscillations induced by the soft modes in the 
infinite-temperature limit. It is also this observation, which restricts the reliability of the result. Nonetheless, 
these results substantiate the non-triviality also of the longitudinal sector and requires further attention. This is especially true for the soft longitudinal mode, where it is yet unclear whether the positivity violation is an artifact of the truncation and if not what is its origin. Known is so far that the extent of such violations depend on the interaction strength \cite{Maas:2004se,Maas:2005rf}. In the case of the transverse gluon, the violation is driven by the vanishing of the propagator in the infrared, and thus a quite robust statement.

\section{Discussion}\label{sdiscuss}

Before discussing further the implications of the results found here on the structure of the Yang-Mills phase diagram, 
it is worthwhile to take a step back to gather and assess the findings. Especially, the reliability of the results 
deserves special attention.

The reliability of the exploratory study of finite temperature effects presented here and thus of the properties of 
the hard modes is a major concern. The success of the truncation scheme in the infinite-temperature limit is based on 
its exactness in the ultraviolet and its presumed exactness in the infrared. Thus the consequences of the deficiencies 
at intermediate momenta are strongly constrained. This constraint is lost by cutting off the Matsubara sums. As a 
consequence, the ultraviolet properties of the hard modes are incorrect and a rooting in perturbation theory is 
prevented. This amplifies the second problem. The hard modes do not reach into the 4-dimensional infrared region due 
to their effective mass. Therefore the advantages of the truncation scheme are lost to a large extent and the 
equations for the hard modes become unreliable.

Nonetheless, the presented results are relevant. At sufficiently large temperatures, the truncation-dependent 
self-energies of the hard modes are suppressed by the large effective mass and they are dominated by the 
truncation-independent tree-level terms. Thus, the truncation artifacts vanish when the temperature goes to infinity. 
The difficult part is then to assess down to which temperature the results are still reliable,
at least qualitatively.

As it is found that the system is only very weakly dependent on temperature, qualitative conclusions can probably be 
drawn even at quite small temperatures. This is due to the fact that the infrared properties of the soft modes are 
nearly independent of the hard modes. In general, the hard modes effectively decouple because of their effective mass, 
which is large compared to $\Lambda_{QCD}$, even at the phase transition. The only exception is the generation of the 
screening mass for the soft longitudinal gluon, which can be found on quite general grounds and thus can be considered 
reliable. This is also supported by the systematic error estimations performed in refs.\cite{Maas:2004se,Maas:2005rf}, 
which do not find qualitative effects but only small quantitative ones.

Although the quantitative results may be subject to change, the qualitative result is expected to be quite reliable. 
Probably down to temperatures within the same order of magnitude as the phase transition, the high temperature phase 
consists of strongly interacting soft modes exhibiting confining properties and inert hard modes. This is supported by 
lattice results, which find that the infinite-temperature limit is effectively reached at about $2T_c$ 
\cite{Cucchieri:2001tw} and the qualitative features of the infinite-temperature limit up to the highest temperatures, 
for which propagators are available, of $6T_c$ \cite{Nakamura:2003pu}. This constitutes the major result found in this 
work.

Concerning the thermodynamic potential, no final conclusion can be drawn. The results indicate that the gross 
thermodynamic properties far away from the phase transition are completely dominated by the hard modes and that a 
Stefan-Boltzmann behavior is reached in the infinite-temperature limit.\quad Considering the crude assumptions made and the 
difficulties encountered, this result is indicative at best. Nonetheless, if it can be substantiated, this would allow 
to understand how a Stefan-Boltzmann-like behavior can emerge from a non-trivial theory. At the same time, the results 
also indicate that, at low temperatures, nonperturbative effects will play a role in the thermodynamics and certainly 
will especially be relevant in the vicinity of the phase transition.

\section{Concluding remarks}\label{ssummary}

Looking at the results found in ref.\ \cite{Maas:2004se,Maas:2005rf} and the solutions at zero \cite{vonSmekal:1997is} 
and small temperatures \cite{Gruter:2004bb}, a coherent picture emerges. Still, this picture is overshadowed by the 
problems of truncation artifacts, especially at finite temperatures.

The main difference between the low-temperature and the high-temperature phase is not primarily one between a strongly interacting and confining 
system and one with only quasi-free particles. The chromoelectric gluons, whose infrared behavior change from 
over-screening to screening, come somewhat close to such a picture, and the hard modes certainly do. The latter become 
free due to the vacuum property of asymptotic freedom and the dependence of their energy on the temperature, thus in 
an expected although not entirely trivial manner. The chromomagnetic gluons remain over-screened in the infrared and 
are thus confined. In the sector transverse to the heat-bath, the results provide strong evidence for the 
Zwanziger-Gribov and/or the Kugo-Ojima scenarios, even at very high temperatures.

Including the observation of super-heating in the low-temperature regime \cite{Gruter:2004bb} and super-cooling at 
high temperatures, the scenario emerging is a chromoelectric phase transition of first order. This phase transition 
connects two different strongly interacting phases of Yang-Mills theory. Especially quantum fluctuations always 
dominate thermal fluctuations and at least part of the gluon spectrum is always confined.

In the vicinity of the phase transition, the consequences of nonperturbative effects are likely also relevant to 
thermodynamic properties, especially to the pressure and the trace anomaly \cite{Zwanziger:2004np}. This underlines 
the importance of nonperturbative effects at least for the temperature range relevant to experiment.

In summary, evidence is found that, in the high-temper\-ature phase of Yang-Mills theory, the elementary excitations 
exhibit quite non-trivial correlations. The soft interactions are non-trivial at all temperatures, and the decoupling 
of the hard modes is evident. This establishes the main features of the high-temperature phase of Yang-Mills theories 
in Landau gauge. The results found here comply with Gribov-Zwanziger or Kugo-Ojima type confinement mechanisms for the 
dimensionally reduced theory. Therefore, the high-temperature phase is definitely nonperturbative. It will be 
interesting to investigate the consequences for quarks and their correlations at high temperature on the one hand and 
for cosmology and the early universe on the other.

\acknowledgement

The authors are grateful to Burghard Gr\"uter for valuable discussions and to Jan M.~Pawlowski for a critical reading of the manuscript and helpful remarks. This work was supported by the BMBF under grant 
numbers 06DA116 and by the Helmholtz association (Virtual Theory Institute VH-VI-041).

\appendix

\section{Running coupling}\label{appcc}

For fixing the coupling there are two options to consider.

If $\mu$ is fixed, $g_3=g^2 T$ grows without limit as $T$ grows. This does not necessarily pose a problem, as the 
infinite-temperature propagators are independent of $g_3$ as long as they are expressed as a function of $p/g_3^2$ 
\cite{Maas:2004se}. Under such circumstances, all momenta below $T$ are effectively infrared, and the nonperturbative 
regime would extend to all momenta. Here this possibility will not be pursued any further\footnote{See 
\cite{Maas:2005rf} for a thorough discussion of this case}.

Alternatively, it is possible to use the limiting prescription $g^2(\mu(T))T=c_\infty$, where $c_\infty$ is an 
arbitrary constant, effectively performing a renormalization group transformation when changing the temperature. This 
fixes $g_3^2$ to be proportional to $\Lambda_{QCD}$, the dynamical scale of QCD \cite{Bohm:yx}, while the 
4-dimensional coupling vanishes like $1/T$. This defines a t'Hooft-like scaling in $T$. Note that $g\to 0$ for 
$T\to\infty$ corresponds to the conventional arguments of a vanishing coupling in the high-temperature phase 
\cite{Blaizot:2001nr}. However, as in the large-$N_c$ limit, a t'Hooft-scaling is performed, generating a well-defined 
theory. Hence, this possibility defines a smooth 3-dimensional limit with a finite 3-dimensional coupling constant. 
Here $c_\infty=1$ is chosen.

Furthermore, it would be useful to give explicit units for the temperature scale. In the vacuum, the scale is fixed 
via a comparison of the running coupling to perturbation theory \cite{Fischer:2002hn}. This is not possible here 
because the Matsubara sum is truncated, and the coupling cannot be calculated reliably for momenta of the order of 
$2\pi NT$.

The simplest procedure is to compare $g_3$ to lattice calculations. There, the effective 3-dimensional coupling is 
found to be $g_3^2(2T_c)$ $/2T_c=$ $2.83$ \cite{Cucchieri:2001tw}. The phase transition temperature is $T_c=269\pm 1$ 
MeV \cite{Karsch:2003jg}. Albeit this will not be exactly $g^2\cdot 2T_c$ due to the truncation, a first approximation is to 
require the same temperature scale at $2T_c$. Using the fixed $g^2T=c_\infty$ prescription, then 
$1=c_\infty/(2T_c\cdot 2.83)$. Since here $c_\infty=1$ in internal units, internal units have to be multiplied by 1.5 
GeV to yield physical units. This temperature scale is used in section \ref{snum}. The ratios of temperatures are 
independent of this prescription.

\section{Subtraction of spurious divergences}\label{appsub}

The spurious divergences in contributions of hard modes due to the integration are dealt with in exactly the same way 
as in 3 dimensions \cite{Maas:2004se} by adjusting the tadpole terms. Each integration kernel $K$ in the longitudinal 
equation is split as
\be
K=K_0+K_D\label{isplit}
\ee
\noindent where $K_0$ is finite and $K_D$ divergent upon integration. Each kernel $K_D$ will be compensated by the 
corresponding tadpole in \pref{fulleqHft}.

A new kind of spurious divergences appear when performing the Matsubara sum. When including more and more Matsubara 
frequencies, it is found that their contribution in the gluon equations \pref{fulleqHft} and \pref{fulleqZft} at 
$|\vec p|<\max(q_0)$ scales as $q_0$, thus behaving as a quadratic divergence. This is an artifact of using a finite 
number of Matsubara frequencies and does not vanish as long as the number of frequencies is finite, no matter how 
large the number. This behavior is spurious and must be removed.

This can be performed by the replacement
\bea
&F(q,q_0)F(p+q,q_0+p_0)\nonumber\\
&\to\left(F(q,q_0)-\frac{1}{Z_3}\right)\left(F(q+p,q_0+p_0)-\frac{1}{Z_3}\right),\label{finite}
\eea
\noindent in all affected loops in the gluon equations. $F$ is a generic dressing function and $Z_3$ its wave-function 
renormalization. 

In the transverse equations all spurious divergences, including the one of the Matsubara sum, are again removed at 
$\zeta=3$. Therefore, the replacement \pref{finite} will only be necessary in those contributions which are 
proportional to $(\zeta-3)$. Hence the integral kernels are split differently than in \pref{isplit} as
\be
K(\zeta)=K_0+(\zeta-3)K_3+K_D(\zeta).\nonumber
\ee
\noindent Here $K_0$ and $K_3$ are finite and independent of $\zeta$, and $K_D$ contains all divergences upon 
integration. The corresponding subtraction is then performed by the replacement
\bea
&&KF(q)F(p+q)\to K_0F(q)F(p+q)\label{transsub}\\
&&+(\zeta-3)K_3\left(F(q)-\frac{1}{Z_3}\right)\left(F(p+q)-\frac{1}{Z_3}\right).\nonumber
\eea

In the soft transverse equation it is necessary in addition to remove tadpole-like structures at $\zeta\neq 3$ in 
$K_3$ \cite{Maas:2005rf}, which alters \pref{transsub} to
\bea
&KF(q)F(p+q)\to K_0F(q)F(p+q)\nonumber\\
&+(\zeta-3)\left(K_3-\frac{1}{p^2}\left(\lim_{p\to 0}p^2K_3\right)\right)\cdot\nonumber\\
&\cdot\left(F(q)-\frac{1}{Z_3}\right)\left(F(p+q)-\frac{1}{Z_3}\right)\nonumber.
\eea

\end{document}